# The FC-rank of a context-free language

A. Carayol

Laboratoire d'Informatique Gaspard-Monge

Université Paris-Est

France

Z. Ésik

Dept. of Computer Science

University of Szeged

Hungary

**Abstract**

We prove that the finite condensation rank (FC-rank) of the lexicographic ordering of a context-free language is strictly less than $\omega^\omega$.

## 1 Introduction

When the alphabet of a language $L$ is linearly ordered, we may equip $L$ with the lexicographic ordering. It is known that every countable linear ordering is isomorphic to the lexicographic ordering of a (prefix) language.

The finite condensation [18] of a linear ordering is obtained by collapsing any two points that are at finite distance of one another. By applying finite condensation iteratively, a fixed point of the condensation map may be reached at some ordinal called the finite condensation rank (FC-rank) of the linear ordering. It is known [18] that for scattered orderings, the FC-rank agrees with the Hausdorff rank.

The lexicographic orderings of regular languages (i.e., the regular linear orderings) were studied in [1, 2, 3, 4, 8, 11, 15, 17, 19]. These linear orderings agree with the leaf orderings of the regular trees, and are all automatic linear orderings as defined in [16]. It follows from results in [15] that all scattered regular linear orderings have finite FC-rank, and in fact all regular linear orderings have finite FC-rank [16]. Moreover, an ordinal is the order type of a regular well-ordering iff it is strictly less than $\omega^\omega$.

The study of the lexicographic orderings of context-free languages (context-free linear orderings) was initiated in [4] and further developed in [5, 6, 12, 13, 14]. It follows from early results in [9] that the lexicographic orderings of deterministic context-free languages are (up to isomorphism) identical to the leaf orderings of the algebraic trees, cf. [5]. In [4], it was shown that every ordinal less than $\omega^{\omega^\omega}$ is the order type of a well-ordered deterministic context-free language, and it was conjectured that a well-ordering is isomorphic to the lexicographic ordering of a context-free language if and only if its order type is less than $\omega^{\omega^\omega}$. This conjecture was confirmed in [5] for deterministic context-free languages, and in [14] for context-free linear orderings. Moreover, it was shown in [6] and [14] that the FC-rank of every scattered deterministic context-free linear ordering and in fact every scattered context-free linear ordering is less than $\omega^\omega$. Since the FC-rank of a well-ordering is less than $\omega^\omega$ exactly when its order type is less than $\omega^{\omega^\omega}$, it follows in conjunction with results proved in [4] that a well-ordering is isomorphic to the lexicographic ordering of a context-free language or deterministic context-free language if and only if its order type is less than $\omega^{\omega^\omega}$. Exactly the same ordinals are the order types of the tree automatic well-orderings, see [10].

In this paper we consider all context-free linear orderings not just the scattered ones. By the above, the FC-rank of a contrext-free linear ordering is at most $\omega^\omega$. Here we prove that it is strictly less than $\omega^\omega$.

For a study of lexicographic orderings of languages generated by *deterministic* higher order grammars we refer to [7].

## 2  Preliminaries

A *linear ordering* [18] $(I, <)$ is a set $I$ equipped with a strict linear order relation $<$. As usual, we will write $x \leq y$ for $x, y \in I$ if $x < y$ or $x = y$. A linear ordering $(I, <)$ is finite or countable if $I$ is. A *morphism* of linear orderings is an order preserving map. Note that every morphism is necessarily injective. When $(I, <)$ and $(J, <')$ are linear orderings such that $I \subseteq J$ and the embedding $I \hookrightarrow J$ is a morphism, we call $(I, <)$ a *subordering* of $(J, <')$. In this case the relation $<$ is the restriction of the relation $<'$ onto $I$ and we usually write just $I$ for $(I, <)$. In particular, an *interval* of $(J, <')$ is a subordering $I$ with the property that whenever $x <' y <' z$ and $x, z \in I$, then $y \in I$. An interval $I$ is called *closed* if there exist $x, y$ in $J$ with $I = \{z : x \leq' z \leq' y\}$. We denote it by $[x, y]$.

An *isomorphism* is a bijective morphism. Isomorphic linear orderings are said to have the same *order type*. The order types of the positive integers $\mathbb{N}$, negative integers $\mathbb{N}_-$, all integers $\mathbb{Z}$, and the rationals $\mathbb{Q}$, ordered as usual, are denoted $\omega$, $\omega^*$, $\zeta$ and $\eta$, respectively. As usual, the finite order types may be identified with the nonnegative integers.

Recall that a linear ordering $(I,<)$ is *dense* if it has at least two elements and for every $x,y \in I$ with $x < y$ there is some $z \in I$ with $x < z < y$. A *quasi-dense* linear ordering is a linear ordering that has a dense subordering, and a *scattered* linear ordering is a linear ordering that is not quasi-dense. For example, $\mathbb{N}$ and $\mathbb{Z}$ are scattered, $\mathbb{Q}$ is dense, and the ordering obtained by replacing each point in $\mathbb{Q}$ with a 2-element linear ordering is quasi-dense but not dense. Clearly, every subordering of a scattered linear ordering is scattered. It is well-known that a linear ordering is quasi-dense iff it has a subordering of order type $\eta$. Moreover, up to isomorphism, there are 4 countable dense linear orderings, the ordering $\mathbb{Q}$ of the rationals possibly equipped with a least or greatest element, or both.

When $(I, <)$ is a linear ordering and for each $i \in I$, $(J_i, <_i)$ is a linear ordering, the *ordered sum*
$$\sum_{i \in I}(J_i, <_i)$$
is the disjoint union $\bigcup_{i \in I}(J_i \times \{i\})$ equipped with the order relation $(x, i) < (y, j)$ iff either $i < j$, or $i = j$ and $x <_i y$. When each $(J_i, <_i)$ is the linear ordering $(J, <')$, we call the ordered sum the *product* of $(I, <)$ and $(J, <')$, denoted $(I, <) \times (J, <')$. Finite ordered sums are also denoted as $(I_1, <_1) + \cdots + (I_n, <_n)$. Since the ordered sum preserves isomorphism, we may also define ordered sums of order types. For example, $1 + \eta + 1$ is the ordered type of the rationals equipped with both a least and a greatest element. It is known that every scattered sum of scattered linear orderings is scattered. This means that if $(I, <)$ is scattered as is each $(J_i, <_i)$, then $\sum_{i \in I}(J_i, <_i)$ is also scattered. A sum over a dense linear ordering $(I, <)$ is referred to as a *dense sum*.

Suppose that $(I, <)$ is a linear ordering. We say that $x, y \in I$ are at a *finite distance* if the intervals $[x, y]$ and $[y, x]$ are both finite (with the convention if $y < x$ then $[x, y]$ is empty).

For each ordinal $\alpha$, we define an equivalence relation $\sim_\alpha^I$ on $I$, together with a linear ordering $(I/\sim_\alpha^I, <_\alpha^I)$ such that for all $x, y$, $x/\sim_\alpha^I <_\alpha^I y/\sim_\alpha^I$ iff $x < y$ and $x$ and $y$ are *not* related by the relation $\sim_\alpha^I$.

The relation $\sim_0^I$ is the identity relation. When $\alpha = \beta + 1$ is a successor ordinal, then for all $x, y \in I$, $x \sim_\alpha^I y$ iff $x/\sim_\beta^I$ and $y/\sim_\beta^I$ are at a finite distance in $(I/\sim_\beta^I, <_\beta^I)$. Suppose now that $\alpha > 0$ is a limit ordinal. Then for all $x, y \in I$, we define $x \sim_\alpha^I y$ iff there exists some $\beta < \alpha$ with $x \sim_\beta^I y$.

By Hausdorff's theorem [18], there is a least ordinal $\alpha = \mathrm{FC}(I, <)$, called the *finite condensation rank* (or *FC-rank*) of $(I, <)$ such that $(I/\sim_\alpha^I, <_\alpha^I)$ is either dense, when $(I, <)$ is quasi-dense, or has at most 1 element, when $(I, <)$ is scattered. In particular, every quasi-dense linear ordering is a dense sum of scattered nonempty linear orderings.

The following facts are known from [18]. The first is stated in Exercice 5.12 on page 82, the second in Lemma 5.14 on page 83, and the third immediately follows from the definition of the FC-rank.

PROPOSITION 2.1  1. *If $(I, <)$ is a linear ordering and $J$ is an interval of $I$, then for all ordinals $\alpha$, $\sim_\alpha^J$ is the restriction of $\sim_\alpha^I$ onto $J$.*

2. *If $(I, <)$ is scattered with a subordering $(J, <)$, then $\mathrm{FC}(J, <) \leq \mathrm{FC}(I, <)$.*

3. *If $(I, <)$ is scattered and $\alpha$ is an ordinal such that $\mathrm{FC}(J, <) \leq \alpha$ for all closed intervals $J \subseteq I$, then $\mathrm{FC}(I, <) \leq \alpha$.*

# 3 Lexicographic orderings

We will consider countable linear orderings that arise as lexicographic orderings of context-free languages. Suppose that $A$ is an alphabet which is linearly ordered by the relation $<$. Then we define a strict partial order $<_s$ on $A^*$ by $u <_s v$ iff $u = xay$ and $v = xbz$ for some $x, y, z \in A^*$ and $a, b \in A$ with $a < b$. We also define $u <_p v$ iff $u$ is a *proper* prefix of $v$, and $u <_\ell v$ iff $u <_s v$ or $u <_p v$. The *lexicographic order* relation $<_\ell$ turns $A^*$ into a linear ordering. In particular, any language $L \subseteq A^*$ gives rise to the linear ordering $(L, <_\ell)$ called the *lexicographic ordering of $L$*. We say that a language $L \subseteq A^*$ is scattered, dense, etc. if its lexicographic ordering has the corresponding property. Moreover, we say that a lexicographic ordering is a *regular* or a *context-free linear ordering* if it is isomorphic to the lexicographic ordering of a regular or context-free language. The FC-rank $\mathrm{FC}(L)$ of a language $L \subseteq A^*$ is the FC-rank of its lexicographic ordering.

We give some examples.

EXAMPLE 3.1 *Consider the alphabet $\mathbf{2} = \{0, 1\}$ ordered by $0 < 1$. Then the lexicographic orderings of the regular languages $1^*0$, $0^*1$, $0^+1 + 1^+0$ are of order type $\omega$, $\omega^*$ and $\zeta$, respectively, so that each is scattered of FC-rank 1. The lexicographic ordering of $(00 + 11)^*01$ is $\eta$. It is thus dense with FC-rank 0. The context-free linear ordering $(\bigcup_{n \geq 0} 1^n 0 (1^*0)^n, <_\ell)$ is of order type $1 + \omega + \omega^2 + \ldots = \omega^\omega$ and has FC-rank $\omega$. The context-free linear orderings $(\bigcup_{n \geq 1} 1^n 0(0(0^+1 + 1^+0) + 10^{<n}), <_\ell)$ and $(\bigcup_{n \geq 1} 1^n 0(0(00+11)^*01 + 1(1^*0)^n), <_\ell)$ with respective order types $\zeta + 1 + \zeta + 2 + \ldots$ and $\eta + \omega + \eta + \omega^2 + \ldots$ have FC-rank 2 and $\omega$, respectively.*

It is known that every countable linear ordering is isomorphic to the lexicographic ordering of a prefix language[1] over the 2-element alphabet $\mathbf{2}$ not containing the empty word $\epsilon$. Similarly, every context-free linear ordering is isomorphic to the lexicographic ordering of a context-free prefix language over $\mathbf{2}$ not containing $\epsilon$. For this, consider an arbitrary context-free language $L$ over an alphabet $A$. Take $\bot$ to be a fresh symbol assumed to be smaller than any symbol of $A$. The lexicographic ordering of the prefix context-free language $L\bot$ is isomorphic to $(L, <_\ell)$. To come back to the binary alphabet $\mathbf{2}$, we then use a standard encoding preserving the order on $A$. Thus, below we may restrict ourselves to such context-free languages which can all be generated by context-free grammars $G = (N, \mathbf{2}, P, S)$ which do not contain useless nonterminals and such that the right-hand side of each production is in $\mathbf{2}^+ N^*$ and does not contain $S$. We say that such grammars are in *weak Greibach normal form* (*weak GNF*). The *height* of such a grammar is the largest integer $n$ such that there is a sequence of nonterminals

$$X_0, \ldots, X_n$$

such that for each $i < n$, $X_{i+1}$ is accessible by a derivation from $X_i$, but $X_i$ is not accessible from $X_{i+1}$.

---
[1]Prefix languages are sometimes called prefix-free.

LEMMA 3.2 *Suppose that $L \subseteq \mathbf{2}^*$ is a context-free prefix language generated by a grammar $G$ in weak GNF of height $n$. Let $u, v \in L$, and suppose that $L_{[u,v]} = \{w \in L : u \leq_s w \leq_s v\}$ is scattered. Then the FC-rank $L_{[u,v]}$ is at most $\omega^n + 1$.*

*Proof.* It suffices to prove this claim when $L_{[u,v]}$ is infinite (and thus $u <_s v$).

Let $G = (N, \mathbf{2}, P, S)$ be a grammar in weak GNF. Let $S'$ be a new nonterminal, and consider all left derivations $S \Rightarrow^* w'p' \Rightarrow wp$, where $w, w' \in \mathbf{2}^*$ with $u <_s w <_s v$ and $w'$ is either a prefix of $u$ or a prefix of $v$, moreover, $p, p' \in N^*$. There can only be finitely many such derivations as $G$ is in weak GNF. For each of these we add a new production $S' \to wp$ to $P'$. In addition we add the two productions $S' \to u$ and $S' \to v$ to obtain a new set of production $P'$. The resulting grammar $G' = (N, \mathbf{2}, P', S')$ generates the scattered language $L_{[u,v]}$. For the inclusion $L(G') \subseteq L_{[u,v]}$, note that a word generated by $G'$ which is not equal to $u$ or $v$ is of the form $wz$ with $u <_s w <_s v$. As by construction $L(G') \subseteq L(G)$, we have $L(G') \subseteq L_{[u,v]}$. In order to prove the opposite inclusion, as $L(G)$ is prefix and $u$ and $v$ belong to $L(G')$, we can restrict our attention to words $x \in L_{[u,v]}$ such that $u <_s x <_s v$. Consider a left-most derivation of $x$ over $G$,

$$S \Rightarrow w_1 p_1 \Rightarrow \cdots \Rightarrow w_n p_n$$

with $w_i \in \mathbf{2}^*$ and $p_i \in N^*$ for $1 \leq i \leq n$, and $w_n = x$ and $p_n = \varepsilon$. Let $\ell$ be the least index such that $w_\ell$ is neither a prefix of $u$ nor a prefix of $v$. Clearly $u <_s w_\ell <_s v$ and hence by the definition of $G'$, $S' \to w_\ell P_\ell$ is a production of $P'$. It follows that $G'$ generates $x$ and hence that $L_{[u,v]} \subseteq L(G')$.

Since the height of (the "reduced part" of) $G'$ is at most $n$, the FC-rank of $L_{[u,v]} = L(G')$ is at most $\omega^n + 1$, by the main result of [14]. □

COROLLARY 3.3 *Suppose that $L \subseteq \mathbf{2}^*$ is a context-free prefix language generated by a grammar $G$ in weak GNF of height $n$. If $L_0 \subseteq L$ is a scattered interval of $L$, then $\mathrm{FC}(L_0) \leq \omega^n + 1$.*

THEOREM 3.4 *The FC-rank of a context-free language is stricly less than $\omega^\omega$.*

*Proof.* Without loss of generality suppose that $L \subseteq \mathbf{2}^*$ is a prefix context-free language. There exists a grammar $G$ in weak GNF of height $n$ generating $L$.

When $L$ is scattered, then the claim holds by [14]. So suppose that $L$ is quasi-dense. Then $L$ can be represented as a dense sum of nonempty scattered linear orderings. This means that $L$ can be partitioned into nonempty scattered intervals $L_i$ indexed by the elements $i$ of a dense countable linear ordering $(D, <)$ such that if $i < j$ in $D$ then $u <_s v$ holds for all $u \in L_i$ and $v \in L_j$. By the previous Corollary and clause (3) of Proposition 2.1, $\mathrm{FC}(L_i) \leq \omega^n + 1$ for each $i$. Thus, $\mathrm{FC}(L) \leq \omega^n + 1$. □

## 4 Conclusion and further research

It has been known that the FC-rank of a regular linear ordering, even automatic linear ordering is finite, cf. [16]. In this note we established the result that the FC-rank of a context-free linear ordering is less than $\omega^\omega$. Our proof method relying on the corresponding result for scattered context-free linear orderings in [14] was the expected one with key ingredient being that the FC-rank of every closed scattered interval in the lexicographic ordering of a context-free prefix language $L$ is bounded by $\omega^n$, where $n$ is a constant depending only on the grammar (in weak GNF) generating $L$.

In [7], it is shown that *scattered* linear orderings generated by certain higher-order deterministic grammars (alias leaf orderings of trees definable by higher-order schemes) all have FC-rank strictly less than $\epsilon_0$, the least ordinal $\alpha$ satisfying the equality $\alpha = \omega^\alpha$. We expect that our methods carry over to prove upper-bounds on the FC-rank of linears orders defined by these higher-order deterministic grammars.

**Acknowledgements**  The authors would like to thank Alexander Kartzow for pointing a mistake in the proof of Lemma 3.2.

# References

[1] S. L. Bloom and C. Choffrut, Long words: the theory of concatenation and omega-power, *Theoretical Computer Science*, 259(2001), 533–548.

[2] S. L. Bloom and Z. Ésik, Deciding whether the frontier of a regular tree is scattered, *Fundamenta Informaticae*, 55(2003), 1–21.

[3] S. L. Bloom and Z. Ésik, The equational theory of regular words, *Information and Computation*, 197(2005), 55–89.

[4] S. L. Bloom and Z. Ésik, Regular and algebraic words and ordinals, in: *CALCO 2007*, Bergen, LNCS 4624, Springer, 2007, 1–15.

[5] S. L. Bloom and Z. Ésik, Algebraic ordinals, *Fundamenta Informaticae*, 99(2010), 383–407.

[6] S. L. Bloom and Z. Ésik, Algebraic linear orderings, *Int. J. Foundations of Computer Science*, 22(2011), 491–515.

[7] L. Braud and A. Carayol, Linear orders in the pushdown hierarchy, in: *ICALP 2010*, LNCS 6199, Springer, 2010, 88–99.

[8] B. Courcelle, Frontiers of infinite trees. *Theoretical Informatics and Applications*, 12(1978), 319–337.

[9] B. Courcelle, Fundamental properties of infinite trees, *Theoretical Computer Science*, 25(1983), 95–169.

[10] C. Delhommé, Automaticité des ordinaux et des graphes homogènes, *C. R. Acad. Sci. Paris, Ser. I*, 339(2004) 5–10.

[11] Z. Ésik, Representing small ordinals by finite automata, in Proc. *12th Workshop Descriptional Complexity of Formal Systems*, Saskatoon, Canada, 2010, EPTCS, vol. 31, 2010, 78–87.

[12] Z. Ésik, An undecidable property of context-free linear orders, *Information Processing Letters*, 111(2010), 107–109.

[13] Z. Ésik, Scattered context-free linear orders, in Proc. *Developments in Language Theory, Milan, 2011*, LNCS 6795, Springer-Verlag, 2011, 216–227.

[14] Z. Ésik and S. Iván, Hausdorff rank of scattered context-free linear orders, in: *LATIN 2012*, Arequipa, Peru, LNCS, to appear in 2012.

[15] S. Heilbrunner, An algorithm for the solution of fixed-point equations for infinite words, *Theoretical Informatics and Applications*, 14(1980), 131–141.

[16] B. Khoussainov, S. Rubin and F. Stephan, Automatic linear orders and trees, *ACM Trans. Comput. Log.*, 6(2005), 625–700.

[17] M. Lohrey and Ch.Mathiessen, Isomorphism of regular words and trees, in: Proc. Automata, Languages and Programming – 38th International Colloquium, ICALP 2011, Zurich, Switzerland, July 4-8, 2011, LNCS 6756, 210–221.

[18] J. G. Rosenstein, *Linear Orderings*, Pure and Applied Mathematics, Vol. 98, Academic Press, 1982.

[19] W. Thomas, On frontiers of regular trees, *Theoretical Informatics and Applications*, vol. 20, 1986, 371–381.